# Machine-learning-based sampling method

# for exploring local energy minima of interstitial species in a crystal

Kazuaki Toyoura[*] and Kansei Kanayama

*Department of Materials Science and Engineering, Kyoto University, Kyoto 606-8501, Japan*

**Abstract**

An efficient machine-learning-based method combined with a conventional local optimization technique has been proposed for exploring local energy minima of interstitial species in a crystal. In the proposed method, an *effective initial point* for local optimization is sampled at each iteration from a given feasible set in the search space. The effective initial point is here defined as the grid point that most likely converges to a new local energy minimum by local optimization and/or is located in the vicinity of the boundaries between energy basins. Specifically, every grid point in the feasible set is classified by the predicted label indicating the local energy minimum that the grid point converges to. The classifier is created and updated at every iteration using the already-known information on the local optimizations at the earlier iterations, which is based on the support vector machine (SVM). The SVM classifier uses our original kernel function designed as reflecting the symmetries of both host crystal and interstitial species. The most distant *unobserved point* on the classification boundaries from the *observed points* is sampled as the next initial point for local optimization. The proposed method is applied to three model cases, i.e., the six-hump camelback function, a proton in strontium zirconate with the orthorhombic perovskite structure, and a water molecule in lanthanum sulfate with the monoclinic structure, to demonstrate the high performance of the proposed method.

* Corresponding author: toyoura.kazuaki.5r@kyoto-u.ac.jp

# I. INTRODUCTION

Point defects and impurities in crystals generally have great influence on materials properties, e.g., electric, optical, and mechanical properties, meaning that adequate control of defect formation leads to improving material properties of interest or adding other materials functions. Therefore, fundamental knowledge on such defects, i.e., defect structures, energetics, and equilibria, is of importance in terms of material design.

First-principles calculations are powerful tools for modeling point defects in recent years [1-7], because of the rapid progress of computer performance and computational techniques to calculate electronic structures in a few decades. In the present day, first-principles point-defect calculations are easy tasks for theorists, and feasible even for experimentalists. The supercell approach under the periodic boundary condition is commonly used [8-13], where our task is only enumerating possible defect structures in a crystal, i.e., possible structures of vacancies, interstitials, substitutional defects, and their complexes.

An important point in the enumeration is that the number of possible defect structures (initial structures for structural optimizations) is largely dependent on the types of defects. In the case of vacancies and substitutional defects, the defect sites are located on the lattice sites of the perfect crystal. Therefore, the number of possible defect structures coincides with the number of crystallographic sites occupied by the defects. By contrast, interstitials have infinite possible positions in principle, although interstitial sites are often limited by exploiting our prior knowledge on solid state physics and chemistry. For example, the tetrahedral and octahedral interstitial sites are of importance in dense inorganic materials. A powerful method for generating possible interstitial sites based on such local structure motifs have recently been proposed, which is called *Interstitial Finding Tool*, InFiT [14]. The algorithm is employed in the *Python Charged Defect Toolkit*, PyCDT [15], and is also implemented in *Python Materials Genomics*, pymatgen [16].

However, such knowledge-based methods have a risk of failure in finding some of interstitial sites in the case of inappropriately-biased prior knowledge. In addition, we occasionally have no prior knowledge on interstitial positions in the case of low-symmetry crystals with relatively-large free space to the size of interstitial species. Such a large space can accommodate not only a single atom but also molecules or atomic groups consisting of several atoms (called polyatomic species hereafter), which increases the degrees of freedom (DOF) of the search space due to the additional three DOF for the rotation of polyatomic species. Hence, higher computational cost could be required for some interstitials than those for vacancies and substitutional defects.

For exploring interstitial sites without prior knowledge, many global optimization algorithms are currently available [17,18], which have been used for structure prediction of crystals and molecules. The simplest methods are the grid search [19] and the random search [20-22], in which local optimizations (finding a neighboring local minimum from a given initial point) are performed using initial points uniformly sampled in a given search space. These straightforward methods should be successful if we set a sufficiently-higher density of initial points than that of local minima. On the other hand, in high-dimensional search spaces, we cannot sample enough initial points for the huge search space, as is often the case with structure prediction of crystals and molecules consisting of many atoms. In such cases, more heuristic methods are conventionally used, e.g., basin hopping [23,24], minima hopping [25,26], simulated annealing [27-29], genetic algorithm [30,31], and metadynamics [32], which are also combined with local optimizations to accelerate the global optimization.

Many global optimization methods are thus available, from which we have to choose a suitable method for the current problem. In the present study, exploration of interstitial sites for polyatomic species in a host crystal is focused on, where both the host crystal and the interstitial species are not accompanied by a drastic change in structure. It is necessary to find not only the global energy minimum (the most stable site) but also other local energy minima (metastable sites) in the search space, because metastable sites could have comparable potential energies to that at the most stable site. The search space has six dimensions at most (three in the translation and three in the rotation), which is expected to have a few dozen of local energy minima at most. In light of the characteristics of the current problem with the relatively small search space, the straightforward methods such as the grid search and the random search with local optimizations seem feasible, which have advantage of uniformly covering the entire search space. In most of more heuristic methods, the search space is explored un-uniformly, in which it could take a great number of steps to escape from a deep energy basin, leading to insufficient exploration in the search space. In addition, such heuristic methods generally have more tuning parameters than the grid search and the random search, implying the necessity of prior knowledge on the potential energy landscape of interstitial species in a given host crystal.

In the present study, a sampling method for initial points of local optimizations has been proposed to explore all local energy minima of interstitial species in a crystal, in which the sampling strategy is made more efficient than those of the grid search and the random search without loss of their simplicity and generality. Prior to explaining the sampling strategy, a few ideal sampling methods are introduced. Figure 1(a) shows the most ideal sampling in the case of a one-dimensional (1D) search space. A single initial point is sampled in each energy basin with a single local minimum, meaning that the minimal number of local optimizations ideally coincides with the number of local energy minima. This is the most efficient sampling, but the energy landscape is practically unknown in advance. In most cases, even the number of local energy minima is not predictable. Therefore, only the information obtained in the ideal sampling is not sufficient to judge when the initial-point sampling for local optimization should be terminated. Note that the trajectories of local optimizations roughly let us know a part of energy landscape, i.e., the sections indicated by double arrows in the figure. Figure 1(b) shows the practically-ideal sampling, in which the trajectories of local optimizations almost cover the entire search space. The initial points are sampled from the vicinity of the boundaries between adjacent basins. In the case of a 1D search space, the number of local optimizations is ideally twice the number of local energy minima. The sampling strategy can be effective also in a higher-dimensional search space to cover the entire energy landscape with the minimal computational cost, as shown in Fig. 1(c) in a 2D search space.

The proposed method for exploring interstitial sites in a crystal is based on the above concept. Due to the unknown energy landscape, we do not know where the local energy minima and the boundaries of energy basins are located in the search space. The next initial point for local optimization at each iteration is here defined as the grid point that most likely converges to a new local energy minimum and/or is located in the vicinity of the boundaries between adjacent energy basins, which is determined by exploiting the already-known information at the earlier iterations. Specifically, all grid points adjacent to the trajectories of local optimizations at the earlier iterations, called *observed points* hereafter, are classified according to the converged local minima. Using the classification as a training data set, a classifier is created on the basis of the support vector machine (SVM) with a kernel [33,34], which estimates the classification boundaries as the decision boundaries and margins. The next initial point to be sampled is here defined as the most distant *unobserved point* in the margins from the observed points. The kernel function is designed on the basis of the periodic kernel as reflecting the symmetries of crystals and interstitial species. The performance of the proposed method is demonstrated using three model systems, i.e., the six-hump camelback function (a 2D test function for global optimization) [35], a proton in strontium zirconate with the orthorhombic perovskite structure ($o$-$SrZrO_3$) [36,37], and a water molecule in lanthanum sulfate with the monoclinic structure ($m$-$La_2(SO_4)_3$) [38-40]. The random search is here used as a reference for the performance comparison, to clearly demonstrate the improvement of the sampling strategy for initial points of local optimizations. The local structure motif method [14] is also used as another reference in the case of proton sites in $o$-$SrZrO_3$, where the usefulness and limitations of the knowledge exploitation are discussed.

## II. PROPOSED METHOD

The outline of the proposed method is shown as the pseudocode in Table 1. In this method, three fundamental sets are defined, i.e., the set of unobserved points, $X_{unobs}$, the set of observed points, $X_{obs}$, and the set $L$ of the sets $L_i$ containing all observed points converging to a local minimum $i$. In the initializing process, $X_{unobs}$ is equal to a given feasible set $X$ in the search space ($X_{unobs} = X$), while $X_{obs}$ and $L$ are null ($X_{obs} = \emptyset$, $L = \emptyset$). In the main loop, an initial point for local optimization is randomly sampled from $X_{unobs}$ while the number of elements in the set $L$, $n_L$, i.e., the number of found local minima, is smaller than two ($n_L < 2$). A local optimization is performed from the sampled initial point, to store all grid points adjacent to the trajectory of the local optimization in the set $A$. $i^{(A)}$ is the index of the local minimum found by the local optimization. If $L$ already includes $L_{i^{(A)}}$, $A$ is added to $L_{i^{(A)}}$ ($L_{i^{(A)}} \leftarrow L_{i^{(A)}} \cup A$). Otherwise, $A$ is added to $L$ as a new element ($L \leftarrow L \cup \{A\}$). The sets of $X_{unobs}$ and $X_{obs}$ are also updated, i.e., $X_{unobs} \leftarrow X_{unobs} \setminus A$ and $X_{obs} \leftarrow X_{obs} \cup A$. Once $n_L \geq 2$, the machine-learning-based sampling method is employed, which samples an unobserved point that most likely converges to a new local minimum by local optimization and/or is located in the vicinity of the boundaries between basins. Specifically, an SVM classifier is created on the basis of the current information of the set $L$. The most distant unobserved point from the observed points is then sampled from the margin of the SVM. This loop is iterated until the distance of any unobserved point in the margin from the first-nearest-neighbor (1NN) observed point ($d_{min}$) is less than a given threshold ($d_{th}$).

In the following subsections, (A) the definition of the feasible sets in various search spaces, (B) the SVM classifier using our original kernel function, and (C) the computational conditions of local optimizations in the present study are individually described in details.

### A. Definition of the feasible sets

In the current problem, not only single-atom species but polyatomic species are also considered as interstitials. Therefore, the feasible set $X$ is generally defined in a 6D search space, i.e., three dimensions for the translation and the other three for the rotation.

As for the three dimensions for the translation, it is conventionally expressed by the fractional coordinates of the interstitial position along the lattice vectors $a$, $b$, and $c$, i.e., $\mathbf{x}_{trans} = (x_a, x_b, x_c)^T$. In the simple case that a crystal of interest only has the translational symmetry along the lattice vectors, the search space is the entire unit cell, leading to the following search space; $0 \leq x_i < 1$ ($i = a, b, c$). If the crystal has additional symmetries such as rotation, mirror, and inverse symmetries, the search space is reduced down to a smaller unit called *asymmetric unit*. Taking $o$-SrZrO$_3$ as an example, the asymmetric unit is defined as $0 \leq x_a < 1$, $0 \leq x_b < 0.5$, and $0 \leq x_c \leq 0.25$ (blue region in Fig. 2(a) left), reflecting the space group of *Pbnm* (62). The volume of asymmetric unit is equal to one-eighth of the unit cell volume. Although the feasible set can be defined by continuous variables in principle, it is expressed as discrete variables in the present study, i.e., sufficiently-fine grid points. In the case of a proton in $o$-SrZrO$_3$, a 20×10×8 grid in the asymmetric unit was used for the feasible set. In the case of a water molecule in $m$-La$_2$(SO$_4$)$_3$ with the space group of $B112/b$ (15), a 25×8×8 grid in the asymmetric unit ($0 \leq x_i \leq 0.5$ ($i = a, b, c$)) was used (Fig. 2(b)).

As for the other three dimensions for the rotation of interstitial species, Euler angles or quaternions are used in general [41]. Instead, we here employ the direction of the principal axis of interstitial species expressed in the spherical coordinate, ($\theta$, $\phi_1$), and the rotational angle around the principal axis, $\phi_2$, which makes it easy to reflect the symmetries of crystals and interstitial species. For example, the expression of a water molecule in a crystal with the Cartesian coordinate is shown in Fig. 2(c). In the initial position, $\mathbf{x}_{rot} = (\theta, \phi_1, \phi_2) = (0, 0, 0)$, the principal axis directs towards the $z$-axis, and the vector from a proton (H1) to the other proton (H2) directs towards the $x$-axis. First, the water molecule is rotated around the principal axis by $\phi_2$. Subsequently, the principal axis is tilted in the ($\theta$, $\phi_1$) direction. The expression $\mathbf{x}_{rot} = (\theta, \phi_1, \phi_2)$ corresponds to the final attitude of the water molecule after the two operations. The search space for the rotation is generally $0 \leq \theta \leq \pi$, $0 \leq \phi_1 < 2\pi$, and $0 \leq \phi_2 < 2\pi$, but $0 \leq \phi_2 < \pi$ in the case of a water molecule due to the rotational symmetry $C_2$ around the principal axis.

The feasible sets for the three model cases are summarized in Table 2. The first model case is the six-hump camelback function, which is a 2D test function for global optimization as a function of $x_1$ and $x_2$. The domain of definition is here set to $-2 \leq x_1 \leq 2$ and $-1 \leq x_2 \leq 1$ without the periodic boundary condition, having six local minima in this domain. The feasible set is 3321 grid points on an 81×41 grid. The second model case is a single proton in $o$-SrZrO$_3$, for which two types of search spaces are defined. The first type is defined on the 3D space grid in the asymmetric unit (Fig. 2(a) left). The feasible set is the grid points on a 20×10×8 grid (grid interval: 0.25 ~ 0.3 Å), from which the grid points close to the host atoms are excluded, leading to 921 grid points in total. The second type of search space has lower dimensions by exploiting prior knowledge on protons in oxides, i.e., an OH bond formation in oxides [42-49]. Specifically, a spherical grid around an oxygen ion is introduced with the radius of 1 Å. This is equivalent to $\mathbf{x}_{rot}$ when the OH ion is regarded as an interstitial species. The OH bond direction corresponds to the principal axis, in which any $\phi_2$ denotes the same OH orientation due to the rotational symmetry C∞ of the OH ion. Therefore, the search space has only two dimensions representing the direction of the principal axis ($\theta$, $\phi_1$). Considering the symmetries on the two oxygen sites in $o$-SrZrO$_3$, the search spaces are $0 \leq \theta \leq \pi/2$ and $0 \leq \phi_1 < 2\pi$ around O1 and $0 \leq \theta \leq \pi$ and $0 \leq \phi_1 < 2\pi$ around O2, where the $a$- and $c$-axes correspond to the $x$- and $z$-axes, respectively. The $\theta$ interval was set to $\pi/12$, while the $\phi_1$ interval was adjusted according to the angle $\theta$. Specifically, the $\phi_1$ interval on the equator ($\theta = \pi/2$) was set to $\pi/12$, and it was adjusted as proportional to the circumferential length at each $\theta$. The spherical grid points around the O1 and O2 ions are shown on the right side in Fig. 2(a), where the grid points close to host atoms are excluded from the feasible sets. As a result, the feasible sets around O1 and O2 contains 83 and 146 points, respectively. In the third model case of a water molecule in $m$-La$_2$(SO$_4$)$_3$, a 25×8×8 grid in the asymmetric unit ($0 \leq x_i \leq 0.5$ ($i = a, b, c$)) was used for $\mathbf{x}_{trans}$, and $\theta$, $\phi_1$, and $\phi_2$ intervals for $\mathbf{x}_{rot}$ were set to $\pi/6$. the $\phi_1$ interval was adjusted at each $\theta$ as in the case of the spherical grid of a proton in $o$-SrZrO$_3$. After excluding the grid points close to host atoms, the number of grid points in the feasible set with six dimensions $\mathbf{x} = (x_a, x_b, x_c, \theta, \phi_1, \phi_2)^T$ are 95217 in total.

**B. SVM classifier**

SVM is one of the conventional methods for binary classification. First, we consider the situation that the training data set contains $n$ elements $\{(\mathbf{x}_i, y_i)\}$ ($i = 1, 2, \ldots, n$), where $\mathbf{x}_i$ is an input, i.e., a grid point in the search space, and $y_i$ denotes the label of point $i$. In the simple case (Fig. 3(a)), the training data can be divided into two classes ($y_i \in \{-1, 1\}$) without exception by a hyperplane in the input space, $f(\mathbf{x}) = \mathbf{w}^T\mathbf{x} + w_0$. In the linear SVM, the classification boundary is defined as the hyperplane whose distance from *support vectors* (1NN points in both classes) is maximized. The region between the two hyperplanes through the support vectors in the same class is called *margin*, particularly *hard margin* in this completely-classable case. The $\mathbf{w}$ and $w_0$ of the classification boundary can be obtained by solving the following optimization problem:

$$\min_{\mathbf{w},w_0} \|w\|^2$$

$$\text{s.t. } y_i(\mathbf{w}^T\mathbf{x}_i + w_0) \geq 1 \ (i = 1, 2, \cdots, n). \quad (1)$$

If permitting some exception for the classification (Fig. 3(b)), the optimization problem becomes as follows:

$$\min_{\mathbf{w},w_0,\xi} \frac{1}{2}\|w\|^2 + C_0 \sum_i \xi_i$$

$$\text{s.t. } y_i(\mathbf{w}^T\mathbf{x}_i + w_0) \geq 1 - \xi_i, \ \xi_i \geq 0 \ (i = 1, 2, \cdots, n), \quad (2)$$

where $\boldsymbol{\xi} = (\xi_1, \xi_2, \ldots, \xi_n)^T$ is the slack variable vector, and $\xi_i > 1$ indicates that the $i^{th}$ training data ($x_i$, $y_i$) is classified by mistake. The second term of the objective function increases with misclassification, and $C_0$ is the regularization parameter to control the degree of misclassification. This optimization problem (primary problem) can be rewritten by the following equivalent problem (dual problem):

$$\max_{\boldsymbol{\alpha}} -\frac{1}{2}\sum_{i,j} \alpha_i \alpha_j y_i y_j \mathbf{x}_i^T \mathbf{x}_j + \sum_i \alpha_i \ (i,j = 1, 2, \cdots, n)$$

$$\text{s.t.} \sum_i \alpha_i y_i = 0, \ 0 \leq \alpha_i \leq C_0 \ (i = 1, 2, \cdots, n), \tag{3}$$

where $\boldsymbol{\alpha} = (\alpha_1, \alpha_2, \ldots, \alpha_n)^T$ are the dual variables. Hence, the classification problem by the SVM reduces to the linearly-constrained convex quadratic optimization problem.

The input space **x** (search space) is often mapped to a higher-dimensional feature space $\Phi(\mathbf{x})$, which enables nonlinear classification in the input space. For classification in the feature space, the inner product of input vectors $\mathbf{x}_i^T \mathbf{x}_j$ in Eq. (3) is replaced by the inner product of feature vectors $\Phi(\mathbf{x}_i)^T \Phi(\mathbf{x}_j)$. Furthermore, $\Phi(\mathbf{x}_i)^T \Phi(\mathbf{x}_j)$ is replaced by the so-called *kernel function* $k(\mathbf{x}_i,\mathbf{x}_j)$, so that we can avoid explicit treatment of variables in the feature space (*kernel trick*). The optimization problem is finally rewritten by

$$\max_{\boldsymbol{\alpha}} -\frac{1}{2}\sum_{i,j} \alpha_i \alpha_j y_i y_j k(\mathbf{x}_i, \mathbf{x}_j) + \sum_i \alpha_i \ (i,j = 1, 2, \cdots, n)$$

$$\text{s.t.} \sum_i \alpha_i y_i = 0, \ 0 \leq \alpha_i \leq C_0 \ (i = 1, 2, \cdots, n). \tag{4}$$

The kernel function was designed as suitable for the symmetries of crystals and interstitial species in the present study. For reflecting the periodicities of crystals and rotational angles, the kernel functions for the translations and rotations of interstitial species are defined in an analogous form to the periodic kernel function as follows:

$$k_0(\mathbf{x}_i, \mathbf{x}_j) = \prod_{\beta=2\pi x_a, 2\pi x_b, 2\pi x_c, \gamma, \phi_2} \exp[C_\beta(\cos(\Delta\beta_{ij}) - 1)], \tag{5}$$

where $\gamma$ denotes the principal-axis direction ($\theta$, $\phi_1$), $\Delta\beta_{ij}$ is the difference in $\beta$ between grid points $i$ and $j$, and $C_\beta$ is the tuning parameter for each $\beta$. The tuning parameters for $x_a$, $x_b$, and $x_c$ were set to $|a|C_1$, $|b|C_1$, and $|c|C_1$ as proportional to the lengths of lattice vectors, respectively, while the other two tuning parameters for the angles $\gamma$ and $\phi_2$ were set to a common value, $C_2$. In addition to the periodicities, the other symmetries of the crystal and the interstitial species should also be taken into account. When $O^{\text{cryst}} = \{O_1^{\text{cryst}}, O_2^{\text{cryst}}, \cdots, O_l^{\text{cryst}}\}$ and $O^{\text{int}} = \{O_1^{\text{int}}, O_2^{\text{int}}, \cdots, O_m^{\text{int}}\}$ denote the sets of all symmetry operations for the crystal and the interstitial species, respectively, a possible kernel function with these symmetries is

$$k(\mathbf{x}_i, \mathbf{x}_j) = \max_{O_l^{\text{cryst}} \in O^{\text{cryst}}, O_m^{\text{int}} \in O^{\text{int}}} k_0(\mathbf{x}_i, O_l^{\text{cryst}} O_m^{\text{int}} \mathbf{x}_j), \qquad (6)$$

which was employed as the kernel function in the present study. Note that the two tuning parameters $C_1$ and $C_2$ must be determined properly, because the kernel matrix based on Eq. (6) is not always positive semi-definite depending on these parameters. The regularization parameter $C_0$ and the two tuning parameters $C_1$ and $C_2$ were here optimized at every iteration by five-fold cross validation, with careful attention to the positive semi-definiteness of the kernel matrix.

The number of local energy minima of interstitial species in a crystal is generally more than two, meaning that the current problem is multi-class classification problem. The one-versus-rest strategy [50] was employed in the present study. The candidates of the next initial point for local optimization was defined as all the grid points located in at least one of the multiple margins.

## C. Computational conditions in local optimization

All calculations for the potential energies of interstitial species in a crystal were performed using first-principles calculations on the basis of the projector augmented wave (PAW) method as implemented in the VASP code [51-54]. The 4$s$, 4$p$, and 5$s$ orbitals for Sr, 4$s$, 4$p$, 5$s$ and 4$d$ orbitals for Zr, 5$s$, 5$p$, 6$s$ and 5$d$ orbitals for La, 3$s$ and 3$p$ orbitals for S, 2$s$ and 2$p$ orbitals for O, and 1$s$ orbital for H were treated as valence states in the PAW potentials. The generalized gradient approximation (GGA) parameterized by Perdew, Burke, and Ernzerhof was used for the exchange-correlation term [55]. The plane wave cut-off energy was set to 400 eV. For a proton in $o$-SrZrO$_3$, a supercell consisting of 2×2×2 unit cells was used with a 2×2×2 mesh for the *k*-point sampling. The charge state of the proton in the supercell was set to +1, which was neutralized by a homogeneous background charge. For a water molecule in *m*-La$_2$(SO$_4$)$_3$, a supercell consisting of 1×2×2 unit cells was used with a single *k*-point sampling at the Γ point. The atomic positions were fully optimized until the residual forces of all atoms became less than 0.02 eV/Å. The conjugate gradient (CG) method [56] was employed for the structural optimization (local optimization).

## III. RESULTS & DISCUSSION

### A. Six-hump camelback function

Prior to exploring interstitial sites in crystals, we demonstrate how the proposed method works on a 2D test function for global optimization, i.e., the six-hump camelback function. This test function is given by

$$f(x_1, x_2) = (4 - 2.1x_1^2 + x_1^4/3)x_1^2 + x_1 x_2 + (-4 + 4x_2^2)x_2^2, \tag{7}$$

which is here limited in the range of $-2 \leq x_1 \leq 2$ and $-1 \leq x_2 \leq 1$ (Fig. 4(a)). This function has six local minima listed in Table 3. In contrast to the wide basins around the two global minima (Mins. 1 and 2), the basins around Mins. 5 and 6 are relatively narrow. This suggests that it is difficult to find these two local minima with narrow basins by local optimizations. Due to the non-periodicity of the test function, the radial basis function (RBF) is used for the kernel function, defined as

$$k_{\text{RBF}}(\mathbf{x}_i, \mathbf{x}_j) = \exp\left(-\|\mathbf{x}_i - \mathbf{x}_j\|^2 / 2C^2\right), \tag{8}$$

where *C* is a tuning parameter.

Figures 4(b)-(h) show the sampling profile of a trial by the proposed method. Note that the grid points adjacent to the transit points in local optimizations are also regarded as observed points in the proposed method. The distance threshold $d_{\text{adj}}$ for the definition of adjacent points to the transit

points was set to 0.05, equal to the grid interval. At the first two iterations, the initial points for local optimizations were randomly sampled, and different local minima were found in this trial. Therefore, the machine-learning-based sampling started at iteration 3 on the basis of the SVM classification. A single initial point was sampled at each iteration from the margins in the current classification, which is the most distant point in the SVM margins from the observed points. At earlier iterations, four local minima with relatively wide basins (Mins. 1–4) were found, and the other two local minima with narrow basins (Mins. 5 and 6) were then found by iteration 25. However, we practically have no prior knowledge on the number of local minima, so that the initial-point sampling for local optimizations continued until satisfying a stopping criterion. In the proposed method, the sampling is terminated when all unobserved points in the margins are close to the observed points. The distance threshold $d_{th}$ was set to 0.2 for this test function. In this trial, the initial-point sampling was stopped at iteration 43. As seen in the final state (Fig. 4(h)), the initial points for local optimizations (white diamond symbol) are located near the borders between basins or the domain boundaries of the test function, not in the deep regions of basins. This indicates that the proposed method was successful in line with the reasonable strategy as shown in Fig. 1. The initial points do not always converge to the nearest local minimum by the employed local optimization based on the CG. Even such a complicated classification can be treated by the soft margin SVM with the RBF kernel. Fig. 4(i) shows the predicted labels of all grid points at the final iteration, and the true labels are also shown in Fig. 4(j) for comparison. The complicated classification in Fig. 4(j) is reasonably predicted by the SVM with only the information on the limited observed points. Note that the minority points are properly neglected in the SVM classification, leading to the somewhat rough classification. The accuracy rate is 88 % in this trial.

  The efficiency and efficacy of the proposed method are discussed hereafter by comparison with the random sampling. We performed 100 trials of the proposed method for the test function, to obtain the profile of the average number of found local minima shown in Fig. 5(a). The profile by the random sampling is also shown, which can theoretically be estimated from the $p^{\mathrm{conv}}$ information in Table 3. The profile of the proposed method converges to the true number of local minima around iteration 40, while that of the random sampling approaches asymptotically to the true number without convergence in this range, indicating higher efficiency of the proposed method. The proposed method exhibits less performance than the random sampling at early iterations before iteration 15. This is probably because the proposed method tends to sample grid points on the domain boundary of the search space at the early iterations. $p^{\mathrm{conv}}$ on the domain boundary are 0.38 for Mins. 1 and 2, 0.12 for Mins. 3 and 4, and 0 for Mins. 5 and 6, which are more scattering than those in the whole domain, leading to the less performance. This tendency is expected to disappear for periodic search spaces without such boundaries. Actually, the less performance cannot be seen in exploring interstitial sites in crystals, as shown in the following subsections.

Figures 5(b) and (c) show the box plots denoting the required number of local optimizations for finding a given number of local minima by the proposed method and the random sampling, respectively. The cases of one or two local minima are not shown in the figure, because the proposed method employs the random sampling until two local minima are found. In the proposed method, all the six local minima can be found by 28.7 local optimizations on average, which is only half number of local optimizations in the random sampling (60.1 local optimizations on average). In addition to the higher efficiency, the proposed method has a superior characteristic that the number of local optimizations has relatively small dispersion to that by the random sampling. In worst-case comparison, less than 50 local optimizations were required by the proposed method, while more than 150 local optimizations were necessary in the random sampling.

The stopping criterion in the present method also works effectively for terminating the sampling. The average number of local optimizations until the sampling termination is 43.1. At any trial, all the six local minima were successfully found before the sampling was terminated, which is the advantage of the proposed method over the random sampling without a clear stopping criterion. A stopping criterion for the random sampling is proposed on the basis of Bayesian statistics [57], in which the sampling is terminated only when the predictive number of local minima $\hat{n}_\text{L}$ is close to the number of found local minima $n_\text{L}$. The number of local minima is predicted by the following equation,

$$\hat{n}_\text{L} = \frac{n_\text{L}(n_\text{loop}-1)}{n_\text{loop}-n_\text{L}-2}, \tag{9}$$

where $n_\text{loop}$ is the current number of iterations. If the typical stopping criterion, $\hat{n}_\text{L} - n_\text{L} = 0.5$, was employed, the required number of iterations after finding the six local minima would be 92. Under this criterion, one-thirds trials of the random sampling would fail to find all six local minima, indicating the difficulty in setting the stopping criterion for the random sampling.

**B. Proton sites in *o*-SrZrO$_3$**

The second model system is proton sites in *o*-SrZrO$_3$, in which the performance of the proposed method is demonstrated for two types of feasible sets. The first feasible set is defined on the 3D space grid in the asymmetric unit of the crystal (Fig. 2(a) left), and the other is the 2D spherical grids around two inequivalent oxygen ions, O1 and O2 ions (Fig. 2(a) right). The obtained results in these two cases are individually demonstrated after showing the true proton sites in this crystal.

Figure 6 shows all proton sites (local energy minima) found by exhaustive local optimizations at all grid points. They all reside around oxygen ions with forming an OH bond, as is the case with protons in oxides [42-49]. There are four proton sites per oxygen ion, which are located near the vertical bisector of the two nearest zirconium ions from the oxygen ion. These proton sites are crystallographically inequivalent due to the low symmetry, in contrast to the equivalent proton sites in cubic perovskites [58-60]. The calculated site energies with reference to the most stable H1 site are scattering in the range of 0 – 0.18 eV as listed in Table 4, which coincide with the reported values in the literature [61]. The problem in this model case is, therefore, that all the eight local energy minima are found by as few local optimizations as possible.

For the first feasible set defined as 921 grid points in the asymmetric unit, 100 trials were performed with different initial points randomly sampled at early iterations. Figure 7(a) shows the profile of the average number of found local energy minima, in which the theoretical profile of the random sampling is also shown for comparison. The profile of the proposed method smoothly converges to the total number of local energy minima, which is faster than that of the random sampling. The proposed method exhibits higher performance at any iteration than the random sampling, which is the difference from the case of the six hump camelback function.

Figures 7(b) and (c) show the box plots denoting required number of local optimizations for finding a given number of local energy minima by the proposed method and the random sampling. The required number of local optimizations for finding all the eight local energy minima are in the range of 8 and 19 depending on the trial (average: 10.9). The required number of local optimizations is fewer than that in the six-hump camelback function, in spite of more local minima in this model case. This is due to the smaller scattering in the basin size around each local energy minimum. The probability that an initial point for local optimization converges to a given local minimum, $p^{\text{conv}}$, is shown in Table 4, which is in the range of 0.072 and 0.193 in this model case. The $p^{\text{conv}}$ range is narrower than that in the six hump camelback function ($0.022 \leq p^{\text{conv}} \leq 0.352$), leading to the efficient sampling. Actually, even the random sampling requires fewer local optimizations, 23.0 on average.

An advantage of the proposed method to the random sampling is the small dispersion in the required number of local optimizations. The standard deviation is 2.3, in contrast to the large standard deviation in the random sampling, 10.0. The maximum numbers of iterations in the 100 trials are 19 in the proposed method vs. 61 in the random sampling.

The other advantage is that the proposed method lets us know when the sampling should be terminated. The required iterations until the sampling termination depend on the distance threshold $d_{th}$. The higher $d_{th}$ requires less iterations, i.e., 124 ± 24, 75 ± 12, and 49 ± 4 at $d_{th}$ = 0.3, 0.4, and 0.5, respectively. However, the higher efficiency at higher $d_{th}$ leads to less accuracy in principle. In this case, the final accuracy rates in the 100 trials are 88 ± 3 %, 82 ± 4 %, and 75 ± 5 % at $d_{th}$ = 0.3, 0.4, and 0.5, respectively. Therefore, $d_{th}$ is regarded as a tuning parameter for adjusting the accuracy vs. efficiency trade-offs in the proposed method. Figure 8 shows the final classifications of a trial at $d_{th}$ = 0.3, 0.4, and 0.5 Å, respectively. In comparison with the true classification, all the classifications seem reasonable, meaning that $d_{th}$ = 0.5 is sufficient in this model case.

The efficiency for exploring interstitial sites can be improved if some reasonable prior knowledge is available. The local structure motif method [14] is widely used for generating possible interstitial sites, which is based on the knowledge that the coordination patterns around interstitials resemble basic structural motifs, e.g., tetrahedral and octahedral environments. In $o$-SrZrO$_3$, seven possible interstitial sites were generated by the local structure motif method, and a proton at each site was finally converged to five proton sites (H1 and H5-8) by local optimizations. The origin of the incomplete exploration is the particularity of the proton sites. Proton sites in oxides are located around a single oxygen ion due to the OH bond formation [42-49], which is different from typical interstitial sites located close to the centers of coordination polyhedra in dense inorganic materials. Note that the scope of the local structure motif method is properly described in the literature [14,15], and that relatively small species to interstitial spaces, complex defects, and polyatomic interstitials are beyond the scope.

The second feasible set defined on the spherical grid around an oxygen ion is based on the reasonable prior knowledge of the OH bond formation in oxides. In $o$-SrZrO$_3$, there are two types of oxygen ions (O1 and O2), and the spherical grid around the O1 ion can be additionally reduced to the semi-spherical grid due to the mirror symmetry. Figure 9(a) shows the sampling profile of a trial on the semi-spherical grid around the O1 ion. The grid points in the vicinity of the pole ($\theta$ =

0) are excluded in advance because proton sites are not located within $ZrO_6$ octahedra, leading to 83 grid points around O1. Four local energy minima (H1–H4) are located on the equator ($\theta = \pi/2$). In this trial, all the four local energy minima was found at iteration 5. The stopping criterion is here defined by the threshold of the angle between the two principal axes, $\gamma_{th}$, instead of $d_{th}$. In this trial, the sampling was terminated at iterations 10 and 15 for $\gamma_{th} = \pi/6$ and $\pi/12$, respectively. The final accuracy rates are 84 % and 99 %, respectively. The final accuracy rates and iterations in 100 trials are 85 ± 4 % and 8.7 ± 0.9 at $\gamma_{th} = \pi/6$ vs. 95 ± 3 % and 14.9 ± 1.5 at $\gamma_{th} = \pi/12$, indicating that $\gamma_{th} = \pi/6$ is sufficient as the stopping criterion.

Around the O2 ion, the search space is the whole spherical grid with 146 grid points except the grid points within the $ZrO_6$ octahedra. Two local energy minima are located near the equator ($\theta = \pi/2$), and the other two local energy minima are around the north and south poles ($\theta = 0$ and $\pi$), respectively. Figure 9(b) shows the sampling profile of a trial around the O2 ion, in which all local energy minima were found at iteration 5. Such quick exploration of local energy minima around the O1 and O2 ions is attributed to the comparable basin size around each local energy minimum, i.e., the comparable $p^{conv}$ listed in Table 4. In this trial, the sampling was terminated at iterations 15 and 25, and the final accuracy rates were 91 % and 96 %, at $\gamma_{th} = \pi/6$ and $\pi/12$, respectively. The final accuracies and iterations in 100 trials are 90 ± 3 % and 12.7 ± 1.3 at $\gamma_{th} = \pi/6$ vs. 97 ± 2 % and 23.3 ± 2.5 at $\gamma_{th} = \pi/12$, respectively.

Thus, the exploration in the second feasible set exhibits higher accuracy and efficiency than that in the first one. This indicates that exploitation of reasonable prior knowledge is effective for exploring interstitial sites, if available.

## C. Water molecule in *m*-$La_2(SO_4)_3$

The third model case is *m*-$La_2(SO_4)_3$, which is reported to exhibit rapid and reversible hydration/dehydration reaction [38-40]. The origin of the smooth insertion/desertion of water molecules is the large interstitial space along the *b*-axis. Figure 10(a) shows the crystal structure of *m*-$La_2(SO_4)_3$ and the most stable site of a water molecule (global energy minima) determined in our recent study using first-principles calculations [39], which was explored by exhaustive structural optimizations with a few assumptions. Specifically, we assumed that water molecules prefer the same site as oxygen ions, and that two OH bonds in the $H_2O$ molecule direct towards the neighboring oxygen ion. The most stable sites are located in the large interstitial space along the *b*-axis. The calculated potential barrier of water diffusion along the *b*-axis is 0.8 eV, while that in the other direction in the *bc*-plane is as high as 1.5 eV. The diffusion pathway along the *a*-axis could not be found in the previous study, whose potential barrier is expected to be higher than those in the *bc*-plane. Thus, the one-dimensional water channel along the *b*-axis enables the rapid and reversible hydration/dehydration reaction.

In the present study, local energy minima including the global energy minimum in *m*-$La_2(SO_4)_3$ were explored in the 6D search space by the proposed method. The feasible set contains 95217 grid points in total, which are the candidates of initial points for local optimizations. Due to the huge computational cost, the true classification of the grid points is unknown, and only a single trial was performed by the proposed method. The two thresholds $d_{th}$ and $\gamma_{th}$ for the stopping criteria were set to 0.5 Å and $\pi/6$, respectively. Figure 11(a) shows the total number of found local energy minima as a function of iterations. A new local energy minimum was discovered frequently at early iterations, the discovery rate gradually slowed down, and finally the sampling was terminated at iteration 433. During this trial, 33 local energy minima including the global energy minimum were found in total.

The proposed method successfully found the global energy minimum corresponding to the most stable site reported in the literature [39]. In addition, the other 32 local energy minima (metastable sites) were found, whose site energies with reference to the most stable site are shown in Fig. 11(b). The site energies are scattering in the wide range of more than 2 eV, roughly classified into two groups, i.e., "five lower-energy minima below 0.6 eV" and "the other higher-energy minima above 1.2 eV". Fig. 10(b) shows the five local energy minima including the global energy minimum in the first group. They all in the first group are located in the water channels along the *b*-axis (Fig. 10(c)).

On the other hand, the local energy minima in the second group are located out of the water channels. The lowest energy minimum in the second group (Min. 6) is located between water channels aligned along the *c*-axis. The relative energy to the global minimum is 1.20 eV, which should be the origin of the higher potential barrier of water diffusion along the *c*-axis. The other local energy minima with higher energies are located in the *inter-channel region* shown by red area in Fig. 10(c). The high site energies suggest the difficulty of the water diffusion along the *a*-axis.

Thus, as many as 33 local energy minima of a water molecule were found in this crystal by the feasible computational cost, i.e., 433 local optimizations. This suggests that the proposed method effectively works also for exploring interstitial sites of polyatomic species requiring the 6D search space.

## IV. CONCLUSIONS

A simple machine-learning-based method for exploring local energy minima of interstitial species in a crystal was proposed in the present study, which is combined with a conventional local optimization technique. In the proposed method, a grid point that most likely converges to a new local energy minimum by local optimization and/or is located in the vicinity of the boundaries between adjacent energy basins is sampled as the next initial point for local optimization at each iteration. Specifically, all observed points at the earlier iterations are classified according to the converged local minimum points, and a classifier is created on the basis of the SVM using the current classification as the training data set. The next grid point is the most distant unobserved point in the margins from the observed points. The kernel function was designed on the basis of the periodic kernel as reflecting the symmetries of crystals and interstitial species.

The performance of the proposed method was demonstrated using three model systems, i.e., the six-hump camelback function, a proton in $o$-$SrZrO_3$, and a water molecule in $m$-$La_2(SO_4)_3$. The results in the first two model cases indicate that the proposed method has higher efficiency than the random sampling for finding all local minima. In addition, the proposed method has a clear stopping criterion, which is a great advantage in contrast to no stopping criterion in the random sampling. In 100 trials for both model cases, the proposed method successfully found all the local minima before the sampling was terminated, indicating the reasonable stopping criteria in the proposed method. In the third case of a water molecule in $m$-$La_2(SO_4)_3$, 33 local energy minima including the global energy minimum were found using the proposed method. The global energy minimum coincides with the previously-reported most stable site of water molecules.

**ACKNOWLEDGMENT**

Useful discussions with Prof. Tetsuya Uda are gratefully acknowledged. This work was partially supported by JSPS KAKENHI Grant Numbers 19H05787 and 20K21082.

**REFERENCES**

[1] A. F. Kohan, G. Ceder, D. Morgan, and C. G. Van de Walle, *Phys. Rev. B* **61**, 15019 (2000).

[2] K. Matsunaga, T. Tanaka, T. Yamamoto, and Y Ikuhara, *Phys. Rev. B* **68**, 085110 (2003).

[3] C. G. Van de Walle and J. Neugebauer, *J. Appl. Phys.* **95**, 3851 (2004).

[4] A. Janotti and C. G. Van de Walle, *Phys. Rev. B* **76**, 165202 (2007).

[5] F. Oba, A Togo, I Tanaka, J Paier, and G Kresse, *Phys. Rev. B* **77**, 245202 (2008).

[6] C. Freysoldt, B. Grabowski, T. Hickel, J. Neugebauer, G. Kresse, A. Janotti, and C. G. Van de Walle, *Rev. Mod. Phys.* **86**, 253 (2014).

[7] H. P. Komsa and A. V. Krasheninnikov, *Phys. Rev. B* **91**, 125304 (2015).

[8] G. Makov and M. C. Payne, *Phys. Rev. B* **51**, 4014 (1995).

[9] S. Lany and A. Zunger, *Phys. Rev. B* **78**, 235104 (2008).

[10] S. Lany and A. Zunger, *Modelling Simul. Mater. Sci. Eng.* **17**, 084002 (2009).

[11] C. Freysoldt, J. Neugebauer, and C. G. Van de Walle, *Phys. Rev. Lett.* **102**, 016402 (2009).

[12] H. P. Komsa, T. T. Rantala, and A. Pasquarello, *Phys. Rev. B* **86**, 045112 (2012).

[13] Y. Kumagai and F. Oba, *Phys. Rev. B* **89**, 195205 (2014).

[14] N. E. R. Zimmermann, M. K. Horton, A. Jain, and M. Haranczyk, *Front. Mater.* **4**, 34 (2017).

[15] D. Broberg, B. Medasani, N. E.R. Zimmermann, G. Yu, A. Canning, M. Haranczyk, M. Asta, and G. Hautier, *Comput. Phys. Commun.* **226**, 165 (2018).

[16] S.P. Ong, W.D. Richards, A. Jain, G. Hautier, M. Kocher, S. Cholia, D. Gunter, V.L. Chevrier, K.A. Persson, and G. Ceder, *Comput. Mater. Sci.* **68**, 314 (2013).

[17] E. M. T. Hendrix and B. G.-Tóth, *Introduction to nonlinear and global optimization* (Springer, Berlin, 2010).

[18] A. R. Oganov, *Modern methods of crystal structure prediction* (Wiley-VCH, Weinheim, 2011)

[19] A. Törn and A. Žilinskas, *Global Optimization, Vol. 350 of Lecture Notes in Computer Science* (Springer, Berlin, Heidelberg, 1989).

[20] R. S. Anderssen, *J. Optim. Theory Appl.* **16**, 383 (1975).

[21] W. L. Price, *J. Optim. Theory Appl.* **40**, 333 (1983).

[22] D. Wales, *Energy landscapes with applications to clusters, biomolecules and glasses* (Cambridge University Press, Cambridge, 2003).

[23] D. J. Wales and J. P. K. Doye, *J. Phys. Chem.* **101**, 5111 (1997).

[24] M. Iwamatsu and Y. Okabe, *Chem. Phys. Lett.* **399**, 396 (2004).

[25] S. Goedecker, *J. Chem. Phys.* **120**, 9911 (2004).

[26] S. Goedecker, W. Hellmann, and T. Lenosky, *Phys. Rev. Lett.* **95**, 055501 (2005).

[27] Kirkpatrick, S., Gelatt, C. D. Jr., and Vecchi, M. P. (1983) Science, 220, 671.

[28] V. Czerny, J. Optim. Theory Appl.45, 41 (1985).

[29] J. Pannetier, J. Bassas-Alsina, J. Rodriguez-Carvajal, and V. Caignart, *Nature* **346**, 343 (1990).

[30] R. L. Johnston, *Applications of Evolutionary Computation in Chemistry, Vol. 110 of Structure and Bonding* (Springer, Berlin, Heidelberg, 2004).

[31] A. R. Oganov and C. W. Glass, *J. Chem. Phys.* **124**, 244704 (2006).

[32] A. Laio, and M. Parrinello, Proc. Nat. Aca. Sci. **99**, 12562 (2002).

[33] C. Cortes and V. Vapnik, *Mach. Learn.* **20**, 273 (1995).

[34] B. E. Boser, I. M. Guyon, V. N. Vapnik, *Proceedings of the fifth annual workshop on Computational learning theory*, 144 (1992).

[35] M. Molga and C. Smutnicki, *Test Functions for Optimization Needs* (2005).

[36] T. Yajima, H. Suzuki, T. Yogo, and H. Iwahara, *Solid State Ion.* **51**, 101 (1992).

[37] T. Omata, Y. Noguchi, and S. O. Y. Matsuo, *J. Electrochem. Soc.* **152**, E200 (2005).

[38] N. Hatada, K. Shizume, and T. Uda, *Adv. Mater.* **29**, 1606569 (2017).

[39] K. Toyoura, H. Tai, N. Hatada, K. Shizume, and T. Uda, *J. Mater. Chem. A* **5**, 20188 (2017).

[40] K. Shizume, N. Hatada, K. Toyoura, and T. Uda, *J. Mater. Chem. A* **6**, 24956 (2018).

[41] C. F. F. Karney, *J. Mol. Graph. Model.* **25**, 595 (2007).

[42] K. D. Kreuer, *Annu. Rev. Mater. Res.* **33**, 333 (2003).

[43] H. Fjeld, K. Toyoura, R. Haugsrud, and T. Norby, *Phys. Chem. Chem. Phys.* **12** 10313 (2010).

[44] L. Malavasi, C. A. J. Fisher, and M. S. Islam, *Chem. Soc. Rev.* **39**, 4370 (2010).

[45] K. Toyoura, N. Hatada, Y. Nose, I. Tanaka, and K. Matsunaga, T. Uda, *J. Phys. Chem. C* **116**, 19117 (2012).

[46] N. Hatada, K. Toyoura, T. Onishi, Y. Adachi, and T. Uda, *J. Phys. Chem. C* **118**, 29629 (2014).

[47] K. Kato, K. Toyoura, A. Nakamura, and K. Matsunaga, *J. Phys. Chem. C* **118**, 9377 (2014).

[48] K. Toyoura, A. Nakamura, and K. Matsunaga, *J. Phys. Chem. C* **119**, 8480 (2015).

[49] K. Toyoura, J. Terasaka, A. Nakamura, and K. Matsunaga, *J. Phys. Chem. C* **121**, 1578 (2017).

[50] C. M. Bishop, *Pattern Recognition and Machine Learning* (Springer, 2006).

[51] P. E. Blöchl, *Phys. Rev. B* **50**, 17953 (1994).

[52] G. Kresse and J. Hafner, *Phys. Rev. B* **48**, 13115 (1993).

[53] G. Kresse and J. Furthmüller, *Comput. Mater. Sci.* **6**, 15 (1996).

[54] G. Kresse and D. Joubert, *Phys. Rev. B* **59**, 1758 (1999).

[55] J. P. Perdew, K. Burke, and M. Ernzerhof, *Phys. Rev. Lett.* **77**, 3865 (1996).

[56] M. R. Hestenes and E. Stiefel, *J. Res. Natl. Bur. Stand.* **49**, 2379 (1952).

[57] C. G. E. Boender and A. H. G. Rinnooy Kan, *Math. Program.* **37**, 59 (1987).

[58] K. Toyoura, D. Hirano, A. Seko, M. Shiga, A. Kuwabara, M. Karasuyama, K. Shitara, and I. Takeuchi, *Phys. Rev. B* **93**, 054112 (2016).

[59] K. Kanamori, K. Toyoura, J Honda, K Hattori, A Seko, M Karasuyama, K. Shitara, M. Shiga, A. Kuwabara, and I. Takeuchi, *Phys. Rev. B* **97**, 125124 (2018).

[60] K. Toyoura, T. Fujii, K. Kanamori, and I. Takeuchi, *Phys. Rev. B* **101**, 184117 (2020).

[61] T. S. Bjørheim, A. Løken, and R. Haugsrud, *J. Mater. Chem. A* **4**, 5917 (2016).

**Table and Figure captions**

TABLE 1. Pseudocode of the proposed method for exploring all local energy minima.

TABLE 2. Summary of feasible sets and given thresholds in the three model systems.

TABLE 3. Six local minima in the six hump camelback function. $p^{conv}$ in the last column denotes the probability that a given initial point for local optimization converges to each local minimum.

TABLE 4. Eight local energy minima of a proton in $o$-SrZrO$_3$. $p^{conv}$ denotes the probability that an given initial point for local optimization converges to each local minimum in the space and spherical grids. Note that the summation of $p^{conv}$ in the spherical grid is unity around each of O1 and O2 ions.

FIG. 1. (Color online) (a) The most ideal sampling and (b) the practically-ideal sampling of initial points for local optimizations in a one-dimensional (1D) search space to find all local minima. The double arrows indicate the regions in which the energy landscape is roughly revealed by the local optimizations. (c) The practically-ideal sampling in a 2D search space.

FIG. 2. (Color online) (a) Two types of feasible sets for exploring local energy minima of a proton in $o$-SrZrO$_3$. One is the 3D space grid in the asymmetric unit, and the other is the 2D spherical grids around O1 and O2 ions with the radius of 1 Å. (b) The space grid in the asymmetric unit of $m$-La$_2$(SO$_4$)$_3$. (c) Definition of the rotational coordinates ($\theta$, $\phi_1$, $\phi_2$) of interstitial species in the case of a water molecule. $\theta$ and $\phi_1$ denote the direction of the principal axis in the spherical coordinate, and $\phi_2$ is the rotational angle around the principal axis. ($\theta$, $\phi_1$, $\phi_2$) denotes the final attitude of interstitial species after the initial rotation around the principal axis and the subsequent tilting of the principal axis.

FIG. 3. (Color online) Examples of binary classification by SVM (a) without and (b) with exceptions.

FIG. 4. (Color online) (a) Six-hump camelback function with six local minima (Mins. 1–6). (b)-(h) Sampling profile of a trial by the proposed method, corresponding to iterations 2, 5, 10, 15, 20, 25, and 43, respectively. The open diamonds are the sampled initial points for local optimizations. The black lines with black points denote trajectories of local optimizations at the last several iterations. The colored solid circles are all observed points adjacent to the transit points of local optimizations in the earlier and current iterations, where the color denotes the local minimum that the grid point converged to. (i) The predicted classification at iteration 43, and (j) the true classification after exhaustive local optimizations from all the grid points. The color denotes the local minimum that the grid point converged to, and all observed points at iteration 43 are fringed with black in the two figures.

FIG. 5. (Color online) For the six-hump camelback function, (a) the profile of the average number of found local minima in 100 trials by the proposed method. The theoretical profile by the random sampling is also shown in the figure for comparison. (b)(c) The box plots denoting required number

of local optimizations for finding a given number of local energy minima by the proposed method and the random sampling, respectively.

FIG. 6. (Color online) All local energy minima of a proton in $o$-SrZrO$_3$ found by exhaustive local optimizations from all grid points. There are four local energy minima around each of O1 and O2 ions.

FIG. 7. (Color online) In the case of exploring local energy minima of a proton on the space grid in $o$-SrZrO$_3$, (a) the profile of the average number of found local energy minima, in which the theoretical profile of the random sampling is also shown for comparison. (b)(c) The box plots denoting the required number of local optimizations for finding a given number of local energy minima by the proposed method and the random sampling, respectively.

FIG. 8. (Color online) In the case of exploring local energy minima of a proton on the space grid in $o$-SrZrO$_3$, (a) the true classification, and (b)-(d) the predicted final classifications of a trial at $d_{th}$ = 0.3, 0.4, and 0.5 Å, respectively.

FIG. 9. (Color online) In the case of exploring local energy minima of a proton on the spherical grids in $o$-SrZrO$_3$, the sampling profiles of a trial by the proposed method around (a) O1 and (b) O2 ions. The open diamonds are the initial points for local optimizations. The black lines with black crosses denote trajectories of local optimizations at the last several iterations. The colored solid circles are all observed points adjacent to the transit points of local optimizations in the earlier and current iterations, where the color denotes the local minimum that the grid point converged to. All grid points in the feasible sets are shown by light gray circles.

Fig. 10. (Color online) (a) The reported most stable site of a water molecule in $m$-La$_2$(SO$_4$)$_3$ [39]. (b) The five local energy minima with lower potential energies in the water channel found by the proposed method. The local energy minima are numbered in the order of potential energy, where zero corresponds to the global energy minimum (the most stable site). The water molecules shown by gray scale are the other local energy minima without the global energy minimum (metastable sites). (c) All the crystallographically-equivalent local energy minima with lower potential energies in the crystal. The water molecules shown by black oxygen ions (Min. 5) are located between two water channels aligned along the $c$-axis.

FIG. 11. (Color online) In the case of exploring local energy minima of a water molecule in the 6D search space of $m$-La$_2$(SO$_4$)$_3$, (a) the profile of the number of found local energy minima in a single trial by the proposed method. (b) Energy levels of 33 local energy minima found by the proposed method with reference to the global energy minimum (Min. 0).

TABLE 1. Pseudocode of the proposed method for exploring all local energy minima.

| Algorithm 1 Local Energy Minima Search ($X$, $LO$, $d_{th}$, $d_{adj}$) |
|---|
| **Initialize:** |
|     Set of unobserved points: $X_{unobs} = X$ |
|     Set of observed points: $X_{obs} = \emptyset$ |
|     Set of the sets $L_i$: $L = \emptyset$ |
|     ($L_i$: Set of all observed points converging to a local minimum $i$) |
| **loop:** |
|     Number of elements in $L$: $n_L$ |
|     **if** $n_L \geq 2$ **then** |
|         Create a classifier based on SVM using the current set $L$ |
|         Sample the most distant unobserved point in the margins from the 1NN observed point |
|         $d_{min}$: Distance between the sampled point and the 1NN observed point |
|     **else** |
|         Randomly sample an unobserved point |
|         $d_{min} = \infty$ |
|     **end if** |
|     $d_{th}$: Distance threshold between the sampled point and the 1NN observed point |
|     **if** $d_{min} \leq d_{th}$ **then** |
|         break the loop |
|     **else** |
|         Perform local optimization ($LO$) |
|         Store all adjacent points to the $LO$ trajectory in the set $A$ |
|         ($d_{adj}$: Distance threshold for the definition of adjacent points) |
|         Subtract the set $A$ from $X_{unobs}$, and add the set $A$ to $X_{obs}$ |
|         **if** $L$ includes $L_{i(A)}$ ($i^{(A)}$: index of the local minimum that the points in $A$ converge to) **then** |
|             Add the set $A$ to $L_{i(A)}$ |
|         **else** |
|             Add the set $A$ in $L$ as a new element $L_{i(A)}$ |
|         **end if** |
|     **end if** |
| **end loop** |

TABLE 2. Summary of feasible sets and given thresholds in the three model systems.

| System | Variables | Domain | Grid points | Thresholds ($d_{th}$, $\gamma_{th}$) |
|---|---|---|---|---|
| Six-hump camelback function | $\mathbf{x}_{trans} = (x_1, x_2)$ | $-2 \leq x_1 \leq 2$<br>$-1 \leq x_2 \leq 1$ | 3321 points<br>81×41 grid | $d_{th} = 0.2$ |
| $H^+$ in $o$-SrZrO$_3$ | Space grid<br>$\mathbf{x}_{trans} = (x_a, x_b, x_c)$ | $0 \leq x_a < 1$<br>$0 \leq x_b < 0.5$<br>$0 \leq x_c \leq 0.25$ | 921 points<br>(20×10×8 grid) | $d_{th} = 0.3, 0.4, 0.5$ Å |
|  | Spherical grid around O1<br>$\mathbf{x}_{rot} = (\theta, \phi_1)$ | $0 \leq \theta \leq \pi/2$<br>$0 \leq \phi_1 < 2\pi$ | 83 points<br>($\theta$ & $\phi_1$ intervals*: $\pi/12$) | $\gamma_{th} = \pi/6, \pi/12$ |
|  | Spherical grid around O2<br>$\mathbf{x}_{rot} = (\theta, \phi_1)$ | $0 \leq \theta \leq \pi$<br>$0 \leq \phi_1 < 2\pi$ | 146 points<br>($\theta$ & $\phi_1$ intervals*: $\pi/12$) | $\gamma_{th} = \pi/6, \pi/12$ |
| H$_2$O in $m$-La(SO$_4$)$_3$ | $\mathbf{x}_{trans}$ & $\mathbf{x}_{rot}$<br>$\mathbf{x}_{trans} = (x_a, x_b, x_c)$<br>$\mathbf{x}_{rot} = (\theta, \phi_1, \phi_2)$ | $0 \leq x_a \leq 0.5$<br>$0 \leq x_b \leq 0.5$<br>$0 \leq x_c \leq 0.5$<br>$0 \leq \theta \leq \pi$<br>$0 \leq \phi_1 < 2\pi$<br>$0 \leq \phi_2 < \pi$ | 95217 points<br>(25×8×8 grid for $\mathbf{x}_{trans}$)<br>($\theta$, $\phi_1$, & $\phi_2$ intervals*: $\pi/6$) | $d_{th} = 0.5$ Å<br>$\gamma_{th} = \pi/6$ |

*$\phi_1$ interval is adjusted as proportional to the circumferential length at each $\theta$.

TABLE 3. Six local minima in the six hump camelback function. $p^{conv}$ in the last column denotes the probability that a given initial point for local optimization converges to each local minimum.

| Min. ID | $x_1$ | $x_2$ | $f(x_1, x_2)$ | $p^{conv}$ |
|---|---|---|---|---|
| 1 | -0.0898 | 0.7127 | -1.0316 | 0.35 |
| 2 | 0.0898 | -0.7127 | -1.0316 | 0.35 |
| 3 | -1.7036 | 0.7961 | -0.2155 | 0.13 |
| 4 | 1.7036 | -0.7961 | -0.2155 | 0.13 |
| 5 | -1.6071 | -0.5687 | 2.1043 | 0.02 |
| 6 | 1.6071 | 0.5687 | 2.1043 | 0.02 |

TABLE 4. Eight local energy minima of a proton in $o$-SrZrO$_3$. $p^{conv}$ denotes the probability that an given initial point for local optimization converges to each local minimum in the space and spherical grids. Note that the summation of $p^{conv}$ in the spherical grid is unity around each of O1 and O2 ions.

| Site ID | 1NN O ion | Site energy (eV) | $p^{conv}$ in space grid | $p^{conv}$ in spherical grid |
|---|---|---|---|---|
| H1 | O1 | 0 | 0.09 | 0.22 |
| H2 | O1 | 0.01 | 0.09 | 0.24 |
| H3 | O1 | 0.03 | 0.10 | 0.35 |
| H4 | O1 | 0.18 | 0.07 | 0.19 |
| H5 | O2 | 0.05 | 0.19 | 0.31 |
| H6 | O2 | 0.09 | 0.18 | 0.29 |
| H7 | O2 | 0.08 | 0.14 | 0.21 |
| H8 | O2 | 0.11 | 0.14 | 0.19 |

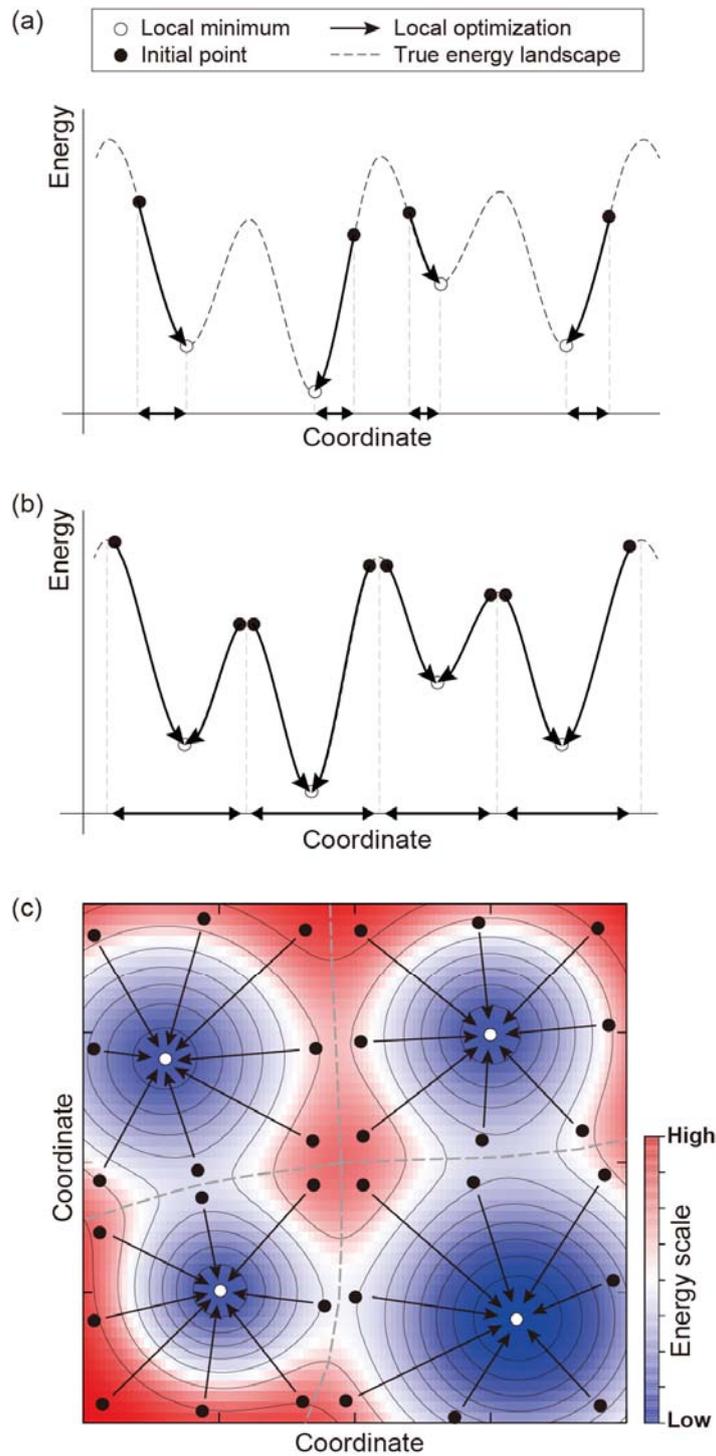

FIG. 1. (Color online) (a) The most ideal sampling and (b) the practically-ideal sampling of initial points for local optimizations in a one-dimensional (1D) search space to find all local minima. The double arrows indicate the regions in which the energy landscape is roughly revealed by the local optimizations. (c) The practically-ideal sampling in a 2D search space.

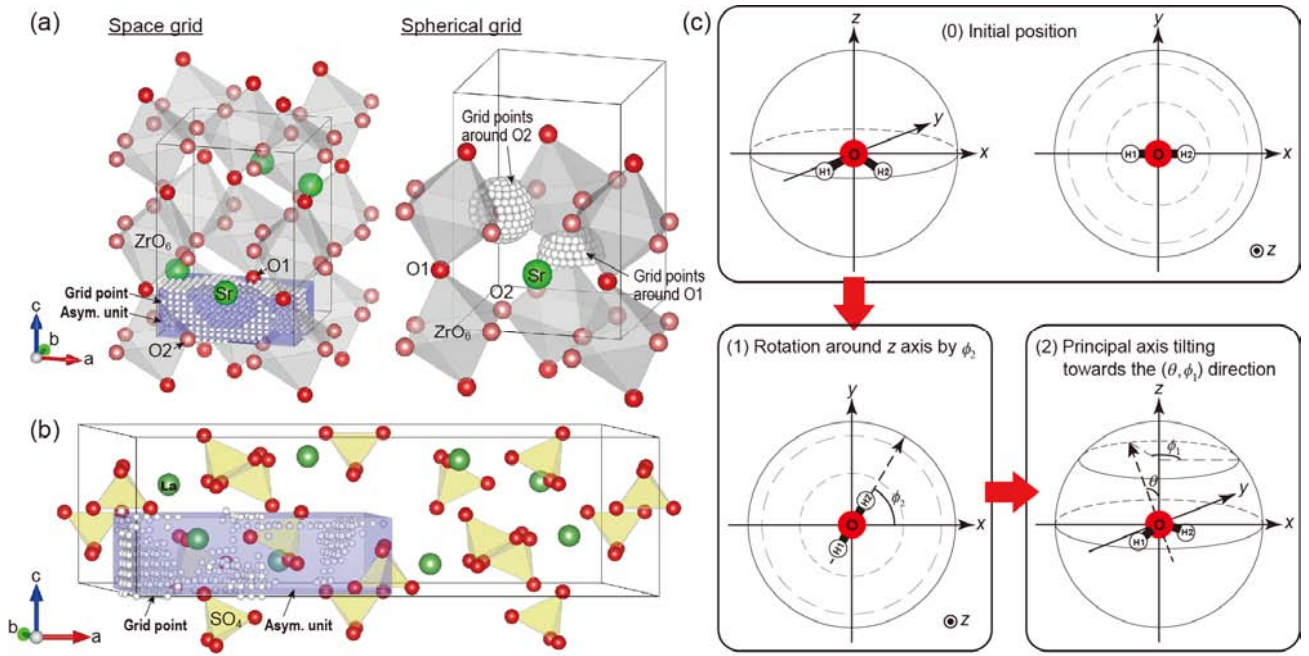

FIG. 2. (Color online) (a) Two types of feasible sets for exploring local energy minima of a proton in $o$-SrZrO$_3$. One is the 3D space grid in the asymmetric unit, and the other is the 2D spherical grids around O1 and O2 ions with the radius of 1 Å. (b) The space grid in the asymmetric unit of $m$-La$_2$(SO$_4$)$_3$. (c) Definition of the rotational coordinates ($\theta$, $\phi_1$, $\phi_2$) of interstitial species in the case of a water molecule. $\theta$ and $\phi_1$ denote the direction of the principal axis in the spherical coordinate, and $\phi_2$ is the rotational angle around the principal axis. ($\theta$, $\phi_1$, $\phi_2$) denotes the final attitude of interstitial species after the initial rotation around the principal axis and the subsequent tilting of the principal axis.

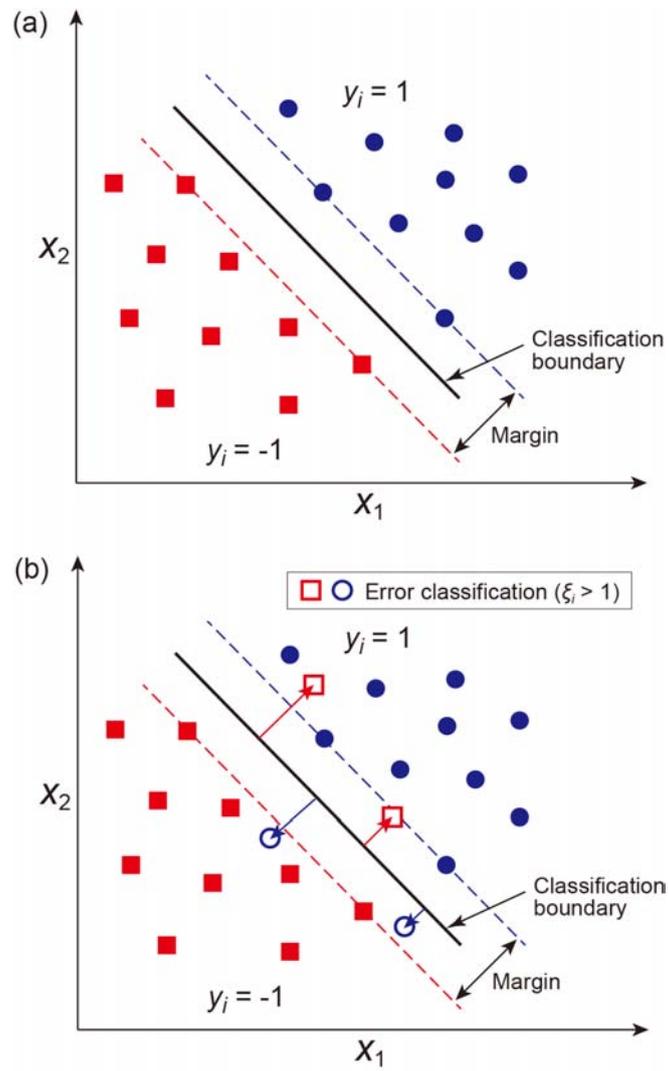

FIG. 3. (Color online) Examples of binary classification by SVM (a) without and (b) with exceptions.

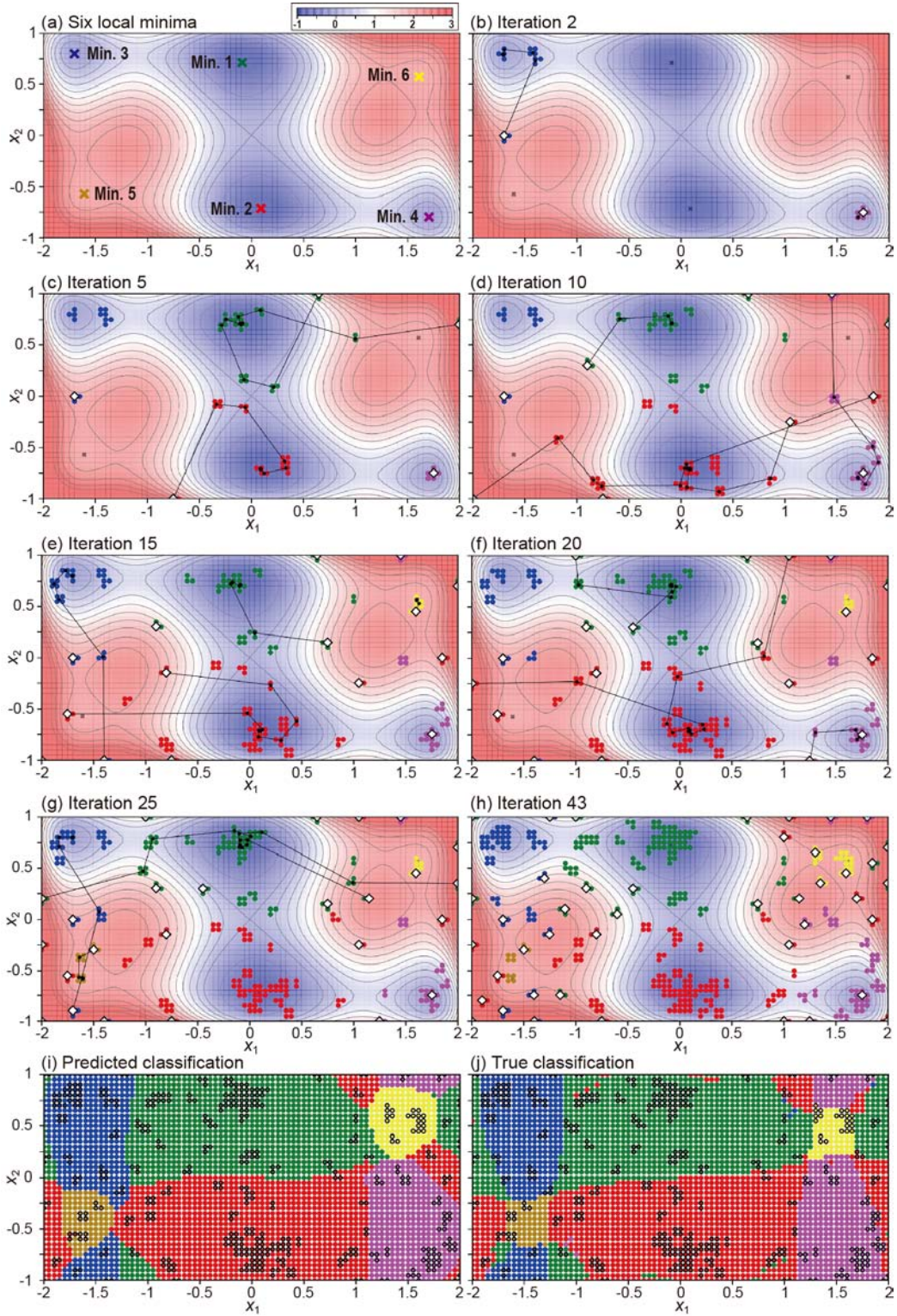

FIG. 4. (Color online) (a) Six-hump camelback function with six local minima (Mins. 1–6). (b)-(h) Sampling profile of a trial by the proposed method, corresponding to iterations 2, 5, 10, 15, 20, 25, and 43, respectively. The open diamonds are the sampled initial points for local optimizations. The black lines with black points denote trajectories of local optimizations at the last several iterations. The colored solid circles are all observed points adjacent to the transit points of local optimizations in the earlier and current iterations, where the color denotes the local minimum that the grid point converged to. (i) The predicted classification at iteration 43, and (j) the true classification after exhaustive local optimizations from all the grid points. The color denotes the local minimum that the grid point converged to, and all observed points at iteration 43 are fringed with black in the two figures.

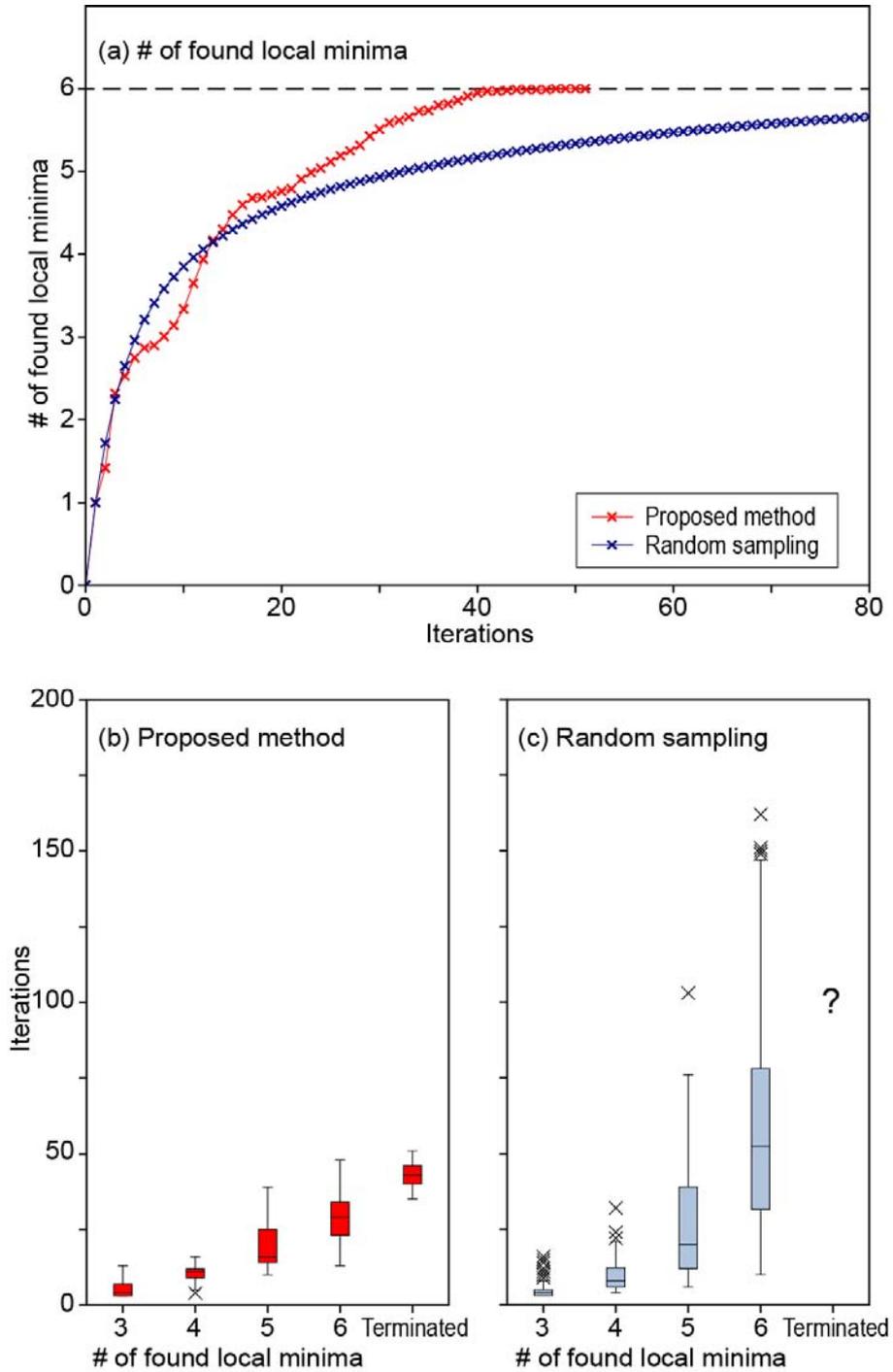

FIG. 5. (Color online) For the six-hump camelback function, (a) the profile of the average number of found local minima in 100 trials by the proposed method. The theoretical profile by the random sampling is also shown in the figure for comparison. (b)(c) The box plots denoting required number of local optimizations for finding a given number of local energy minima by the proposed method and the random sampling, respectively.

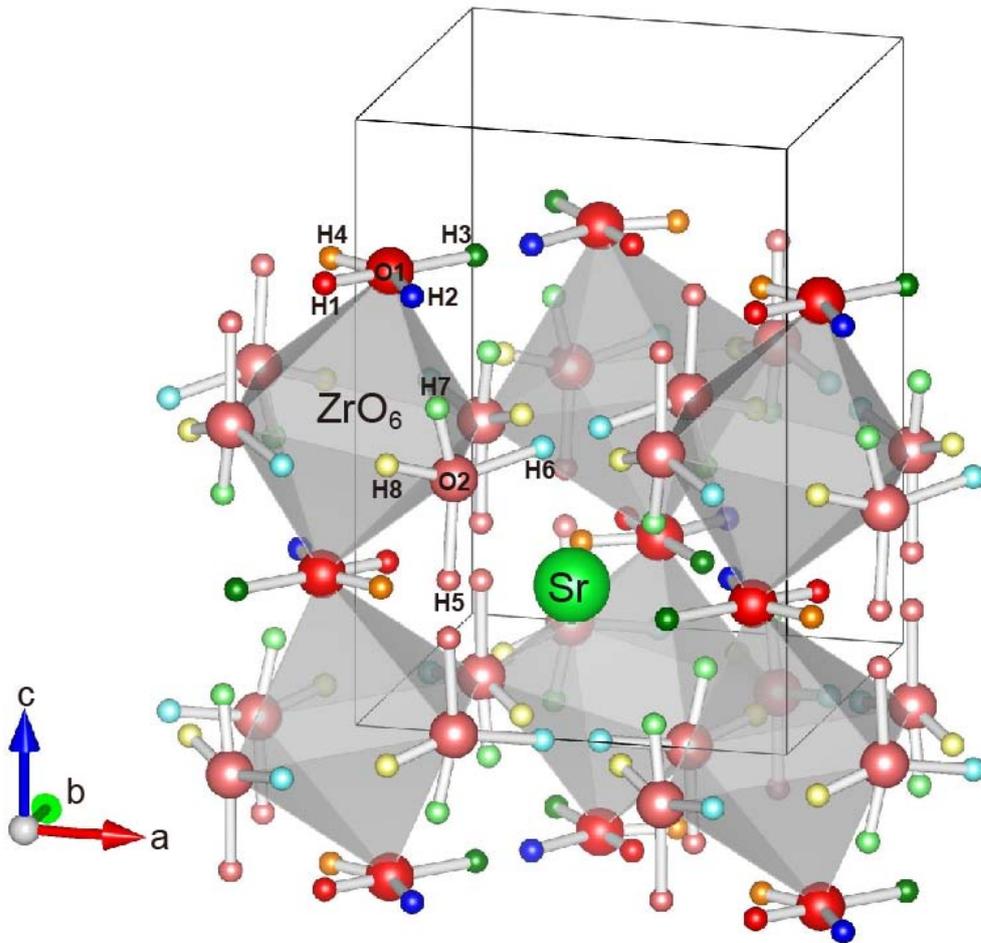

FIG. 6. (Color online) All local energy minima of a proton in $o$-SrZrO$_3$ found by exhaustive local optimizations from all grid points. There are four local energy minima around each of O1 and O2 ions.

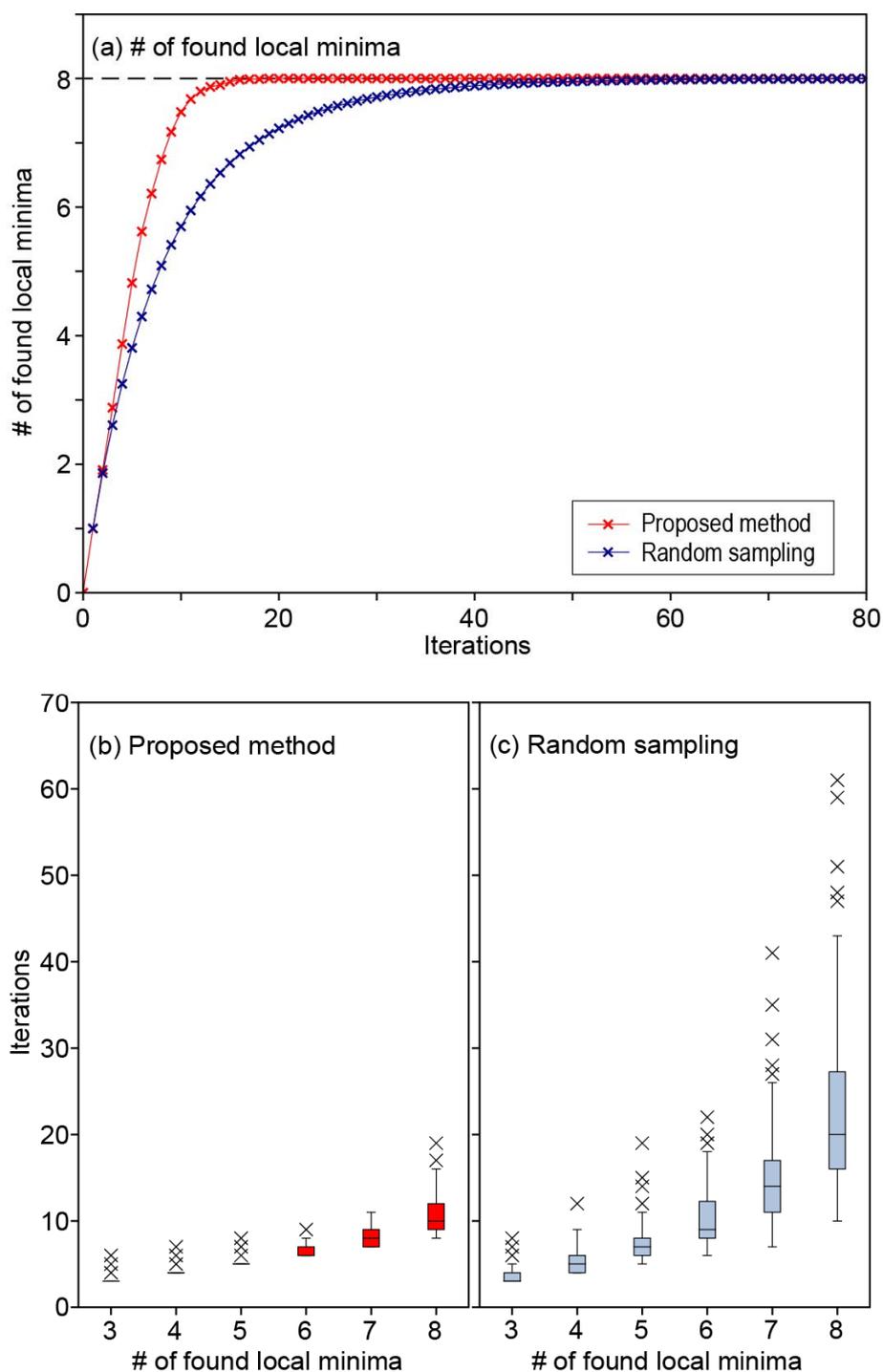

FIG. 7. (Color online) In the case of exploring local energy minima of a proton on the space grid in $o$-SrZrO$_3$, (a) the profile of the average number of found local energy minima, in which the theoretical profile of the random sampling is also shown for comparison. (b)(c) The box plots denoting the required number of local optimizations for finding a given number of local energy minima by the proposed method and the random sampling, respectively.

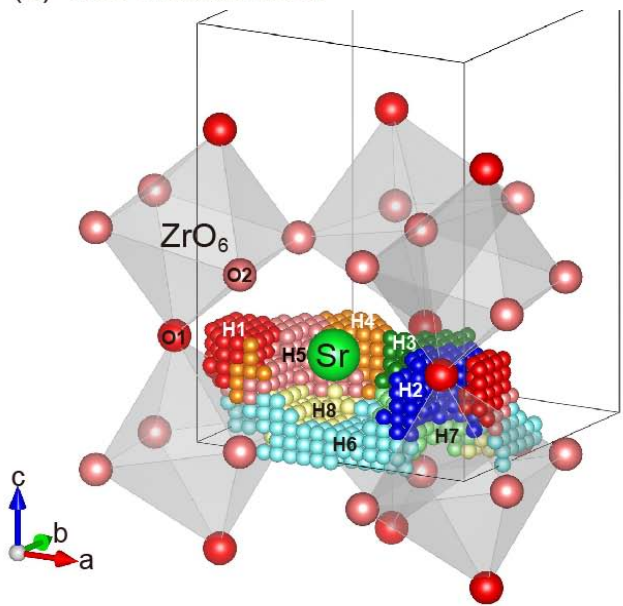
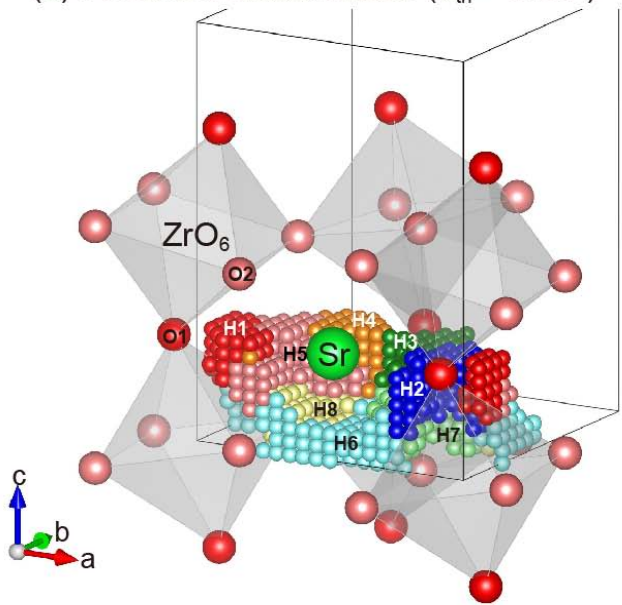
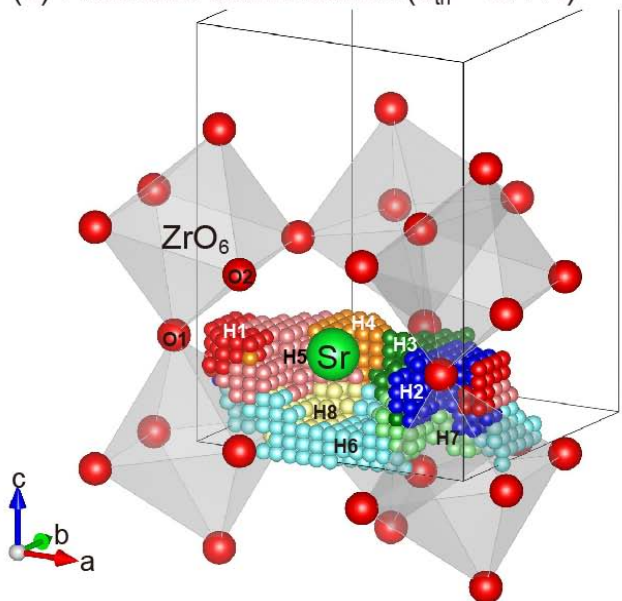
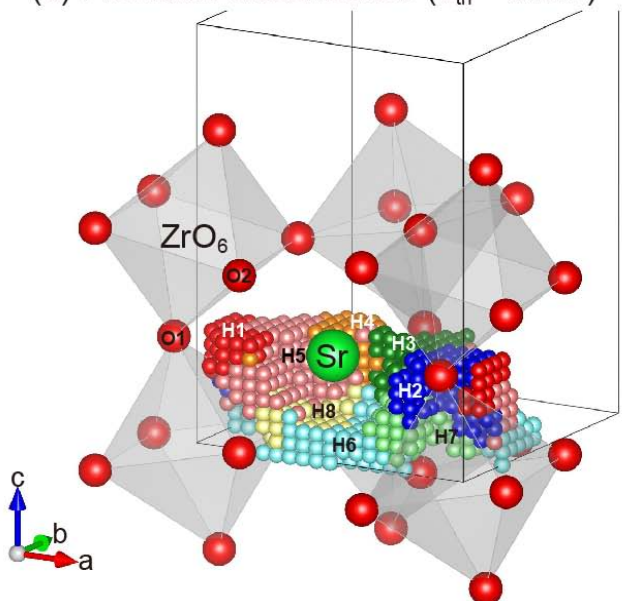

FIG. 8. (Color online) In the case of exploring local energy minima of a proton on the space grid in $o$-SrZrO$_3$, (a) the true classification, and (b)-(d) the predicted final classifications of a trial at $d_{th}$ = 0.3, 0.4, and 0.5 Å, respectively.

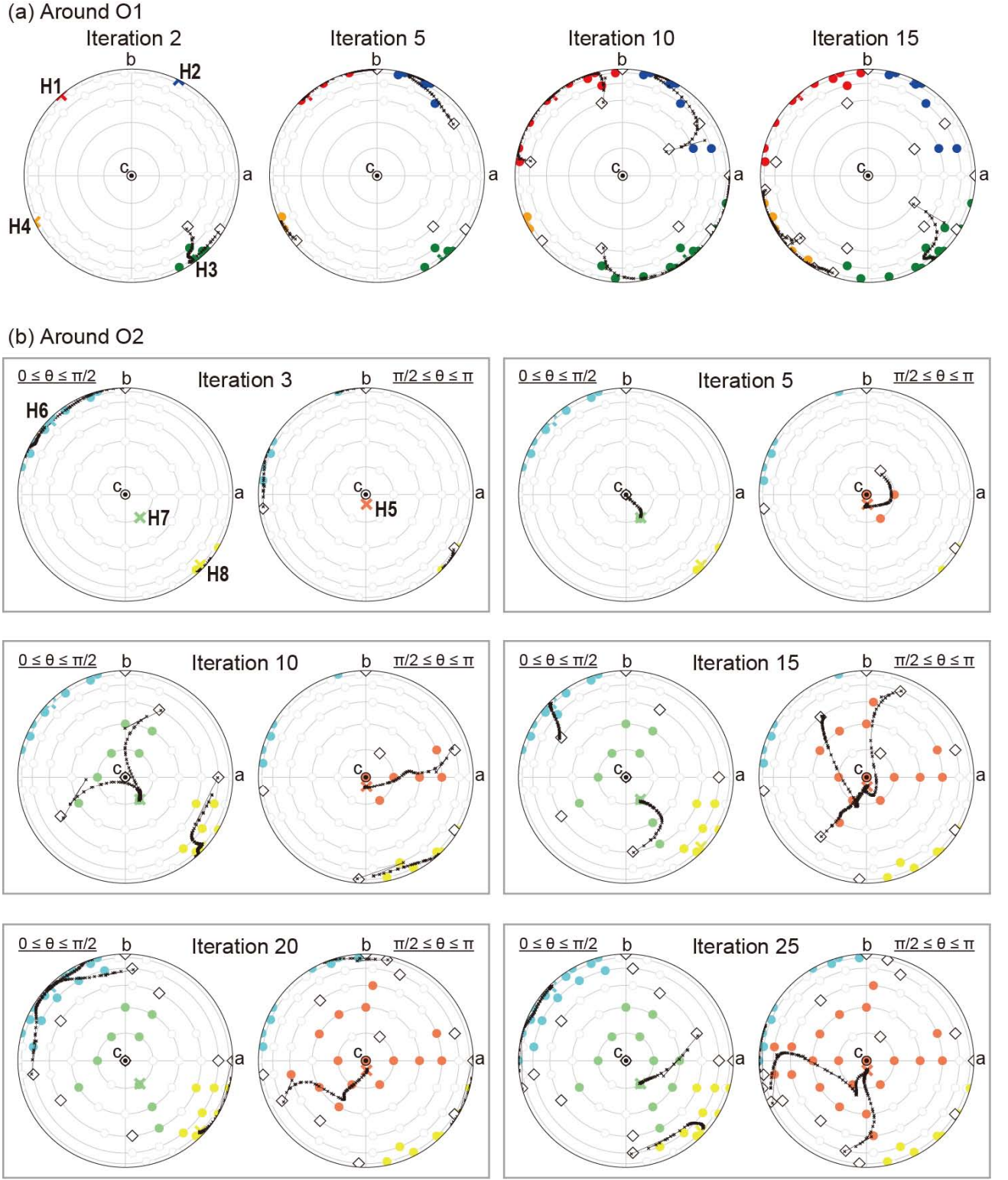

FIG. 9. (Color online) In the case of exploring local energy minima of a proton on the spherical grids in $o$-SrZrO$_3$, the sampling profiles of a trial by the proposed method around (a) O1 and (b) O2 ions. The open diamonds are the initial points for local optimizations. The black lines with black crosses denote trajectories of local optimizations at the last several iterations. The colored solid circles are all observed points adjacent to the transit points of local optimizations in the earlier and current iterations, where the color denotes the local minimum that the grid point converged to. All grid points in the feasible sets are shown by light gray circles.

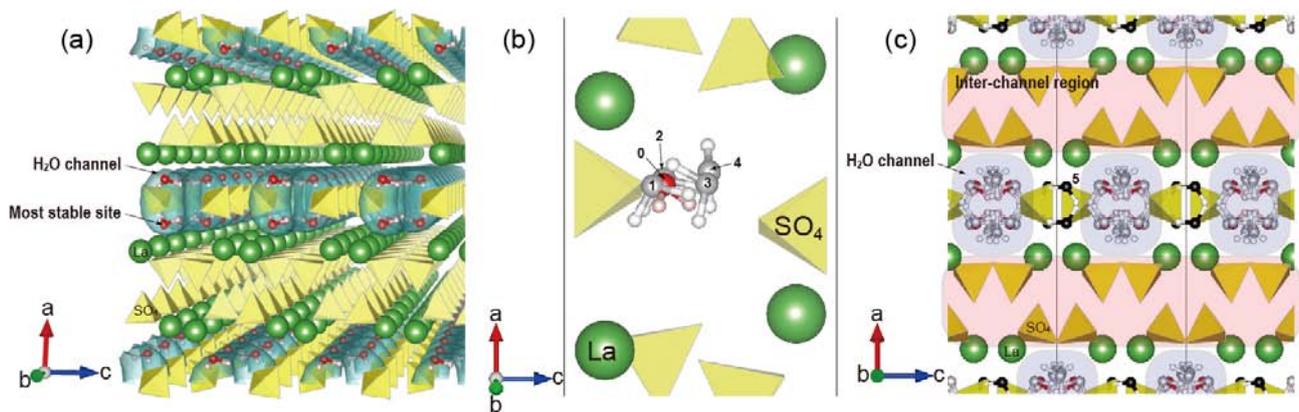

Fig. 10. (Color online) (a) The reported most stable site of a water molecule in $m$-La$_2$(SO$_4$)$_3$ [39]. (b) The five local energy minima with lower potential energies in the water channel found by the proposed method. The local energy minima are numbered in the order of potential energy, where zero corresponds to the global energy minimum (the most stable site). The water molecules shown by gray scale are the other local energy minima without the global energy minimum (metastable sites). (c) All the crystallographically-equivalent local energy minima with lower potential energies in the crystal. The water molecules shown by black oxygen ions (Min. 5) are located between two water channels aligned along the $c$-axis.

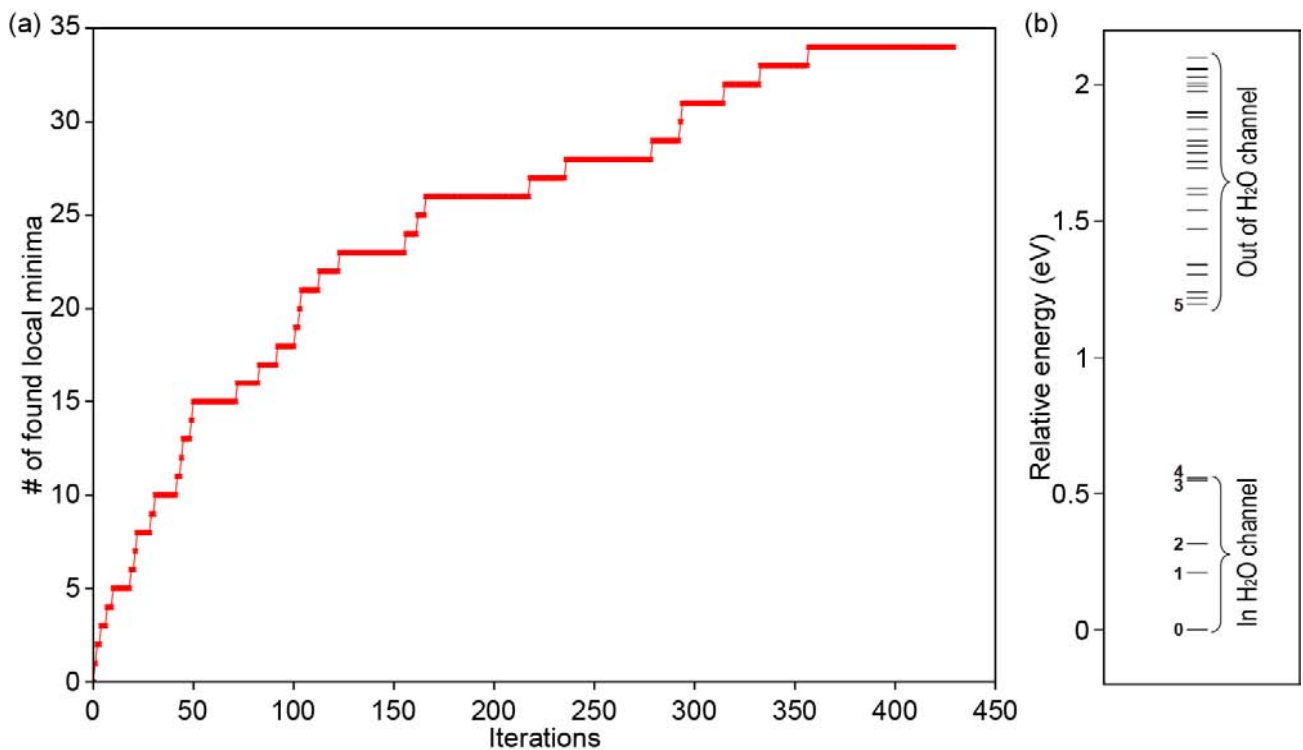

FIG. 11. (Color online) In the case of exploring local energy minima of a water molecule in the 6D search space of $m$-$La_2(SO_4)_3$, (a) the profile of the number of found local energy minima in a single trial by the proposed method. (b) Energy levels of 33 local energy minima found by the proposed method with reference to the global energy minimum (Min. 0).